\documentclass{PoS}

\title{Dark Matter and Leptogenesis in the Inverse Seesaw}

\ShortTitle{Dark Matter and Leptogenesis in the Inverse Seesaw}

\author{\speaker{Fran\c{c}ois-Xavier Josse-Michaux}\thanks{E-mail: fxjossemichaux@gmail.com} \, and Emiliano Molinaro\\
        \\
       Centro de F\'isica Te\'orica de Part\'iculas (CFTP), Instituto Superior T\'ecnico,\\
Technical University of Lisbon, 1049-001 Lisboa, Portugal \\
       } 

\abstract{We study  a seesaw-type extension of the Standard Model
in which the symmetry group is enlarged by a global $U(1)$.  
We introduce  adequate scalar and fermion representations which naturally explain the smallness of neutrino masses. 
With the addition of a viable scalar Dark Matter candidate, an original and successful scenario of leptogenesis emerges.}
\FullConference{XXIst International Europhysics Conference on High Energy Physics\\
          21-27 July 2011\\
          Grenoble, Rhones Alpes France}

\begin{document}
It is well known that leptogenesis in the type I seesaw cannot work at the TeV scale, unless a strong degeneracy exists among the heavy right-handed neutrinos that are introduced to explain the small neutrino masses, cf. eg.~\cite{Davidson:2008bu}.
However, such a degeneracy is quite unnatural in the type I seesaw, and one rather advocates an inverse-seesaw mechanism~\cite{Mohapatra:1986bd}, where the small mass splitting emerges from the breaking of an underlying symmetry, usually related to the accidental lepton number symmetry the Standard Model (SM) exhibits. 
In this proceeding, based on \cite{JosseMichaux:2011ba}, we present a UV-completion of the inverse-seesaw, promoting the accidental symmetry to a global conserved charge and introducing the adequate Higgs representations accounting for non-zero neutrino masses. 
We further elaborate on this model by introducing a Dark Matter (DM) candidate, under the form of a scalar charged under the lepton (-like) charge, and whose very presence supplies an original scenario of leptogenesis, successful down to the TeV scale.
\section{A UV-completion of the inverse-seesaw}
We introduce a global $U(1)$ symmetry conserving a $B-\tilde{L}$ charge. 
$B$ is the standard Baryon number, while $\tilde{L}$ is the Lepton number extended to non-SM particles, thus quarks have charge $+1$ under this symmetry, while SM leptons have charge $-1$. 
The SM Higgs boson $H_{1}$ has zero $B-\tilde{L}$ charge. 
In the inverse-seesaw mechanism, two right-handed neutrinos are introduced with opposite $\tilde{L}$ charges $\tilde{L}(N_{1})=1=-\tilde{L}(N_{2})$. 
We conveniently form a Dirac spinor $N_{D}$ with $\tilde{L}(N_{D})=1$ from both fields $N_{D}\equiv P_{R}\,N_{1}+P_{L}\,N_{2}^c$, which couples to the SM-leptons through a Yukawa interaction $y_{1}\,\overline{N_{D}}\,\tilde{H_{1}}^{\dagger}\,\ell+h.c.\,$ and receives a Dirac mass $M\, \overline{N_{D}}\, N_{D}$.
As shown in \cite{Kersten:2007vk} for example, with only this particle content, SM active-neutrinos are massless. 
The inverse-seesaw completion is obtained by introducing a scalar $H_{3}$, singlet under the SM gauge group but with $\tilde{L}(H_{3})=2$, such that a Yukawa coupling can be written $\alpha H_{3}\,\overline{N_{D}}\,N_{D}^c+h.c.\,$, yielding a Majorana mass once $H_{3}$ acquires a non-zero vacuum expectation value $\propto \alpha\, v_{3}\, \overline{N_{D}}\,N_{D}^c$. 
Yet the model accounts only for one massive light neutrino \cite{ThomasBelen}, so we extend the particle content with a scalar doublet $H_{2}$, with $\tilde{L}(H_{2})=-2$. 
A Yukawa coupling $y_{2}\,\overline{N_{D}^c}\,\tilde{H_{2}}^{\dagger}\,\ell+h.c.\,$ is formed, which provides a Dirac mass term after $H_{2}$ takes a non-zero vev.
The light neutrino mass is given by
\begin{equation} \label{mnu}
\left(m_{\nu}\right)^{ij} = -y_{ 1}^{\lbrace i\,,}\,y_{2}^{j \rbrace }\frac{v_{1}\, v_{2}}{2\,M}+y_{1}^{i}\, y_{1}^{j}\,\alpha^{*}\,v_{3}\,\frac{v_{1}^2}{2\,M^2}+y_{2}^{i}\,y_{2}^{j}\,\alpha\,v_{3}\frac{v_{2}^2}{2\,M^2}\,,\quad v_{i}=\sqrt{2}\,\left\langle H_{i} \right\rangle\,.
\end{equation}
As clear from the equation above, the introduction of $H_{2}$ is mandatory to obtain two-massive light neutrinos, as required by observations. Similarly $(y_{1}^e,y_{1}^\mu,y_{1}^\tau)$ and $(y_{2}^e,y_{2}^\mu,y_{2}^\tau)$ should not be aligned. As for the terms $\propto \alpha$, even if not necessary for neutrino masses, they are crucial for leptogenesis~\cite{JosseMichaux:2011ba}. Furthermore, from the spontaneous breaking of the global $U(1)_{B-\tilde{L}}$ by the non-zero vev of $H_{2}$ and $H_{3}$ emerges a massless Goldstone boson, the Majoron, whose couplings to the SM particles are strongly constrained: this enforces a hierarchy among the Higgs vevs: $v_{2} \ll v_{1,3}$, naturally realized in the model~\cite{JosseMichaux:2011ba}. Therefore, the second term of eq.(\ref{mnu}) dominates over the other terms, and the model effectively realizes an inverse-seesaw for neutrino masses.
\section{A 2-step leptogenesis scenario}
The realization described above explains neutrino masses, but yet fails to account for what makes the seesaw model so attractive: the explanation of the baryon asymmetry of the Universe via the leptogenesis scenario~\cite{lepto}.
To this end, we introduce a third right-handed neutrino, singlet under all conserved charges and of Majorana type, together with a complex scalar field $S$, a SM singlet but with $\tilde{L}(S)=1$. 
The fermion singlet does not participate in neutrino mass generation, as no Higgs representation with $\tilde{L}=\pm 1$ are introduced. 
Several extra couplings are allowed by the symmetry of the model, among which the most relevant ones are a Yukawa coupling $g\,\overline{N_{3}}\,N_{D}\,S^{*}+h.c.$ and a trilinear term $\mu^{\prime \prime} S\,S\,H_{3}+h.c.$. 
Through the exchange of $H_{3}$, a $CP$ asymmetry can be generated in $N_{3}$ decays, and a leptogenesis scenario arises. 
The generation of a baryon asymmetry proceeds in two stages:
\begin{itemize}
\item A first step in which asymmetries in $N_{D}$ and $S$ (and $H_{3}$) are created, by decays of $N_{3}$, inverse decays, scatterings ... . This stage is similar to the standard leptogenesis scenario in the type I seesaw. 
\item A second step during which the asymmetries in $N_{D}$ and $S$ are transferred to the SM leptons, through Yukawa couplings $y_{1}$ or $y_{2}$ driven interactions.
\end{itemize}
The lepton asymmetry produced in the second stage is reprocessed, partially, into a baryon asymmetry via sphaleron interactions. 
Note that the processes responsible for the two steps are/may be in thermal equilibrium at the same temperature, but the generation of an asymmetry in $N_{D}$ is a required prior to the generation of a non-zero lepton asymmetry.

As in the standard leptogenesis scenario, the amount of $N_{D}$ and $S$ asymmetries produce during the first stage depends on the size of the $CP$ asymmetry in $N_{3}$ decays and on the efficiency of leptogenesis, which results form a competition between production and depletion processes. 
In the limit where $N_{3}$ is much heavier than $N_{D}$, the $CP$ asymmetry approximately equals
\begin{equation}
\epsilon_{CP}\simeq -2\times 10^{-6}\,\left(\frac{\mu^{\prime\,\prime}}{1 \,{\rm GeV}}\right)\,\left(\frac{10\,\,{\rm TeV}}{M_3}\right)\,{\rm Im}(\alpha)\,,
\end{equation}
where $M_{3}$ is the Majorana mass of $N_{3}$. 
For non-suppressed $\alpha$ and ratio $\mu^{\prime\,\prime}/M_{3}$, the $CP$ asymmetry is typically large enough. 
Concerning the efficiency of asymmetry production, several processes have to be considered. 
In a rough picture, decays and inverse decays dominate and values of $g\lesssim {\rm few\,}10^{-5}$ are required, for an optimal efficiency. 
The interesting point is that the $CP$ asymmetry does not depend on the coupling $g$ responsible of these washout. 
This is in sharp contrast with the leptogenesis scenario in the type I seesaw, where both depend on the neutrino Yukawa couplings.

The transfer of the asymmetry stored in $N_{D}$ to the SM leptons occurs through the decays of $N_{D}$ and through Higgs-mediated scatterings: all depend on the neutrino Yukawa couplings $y_{1,2}$. 
We show in \cite{JosseMichaux:2011ba} that when neutrino mass constraints are applied, the transfer is efficient and the leptonic asymmetry equals the asymmetry in $N_{D}$ produced in the first stage. 
Thus basically, for neutrino Yukawa large enough, leptogenesis depends only on $\mu^{\prime \prime}$, $\alpha$ and $g$ and is easily successful.
\section{Dark Matter}
\begin{figure}[h!]
\begin{center}
\includegraphics[width=0.6\textwidth]{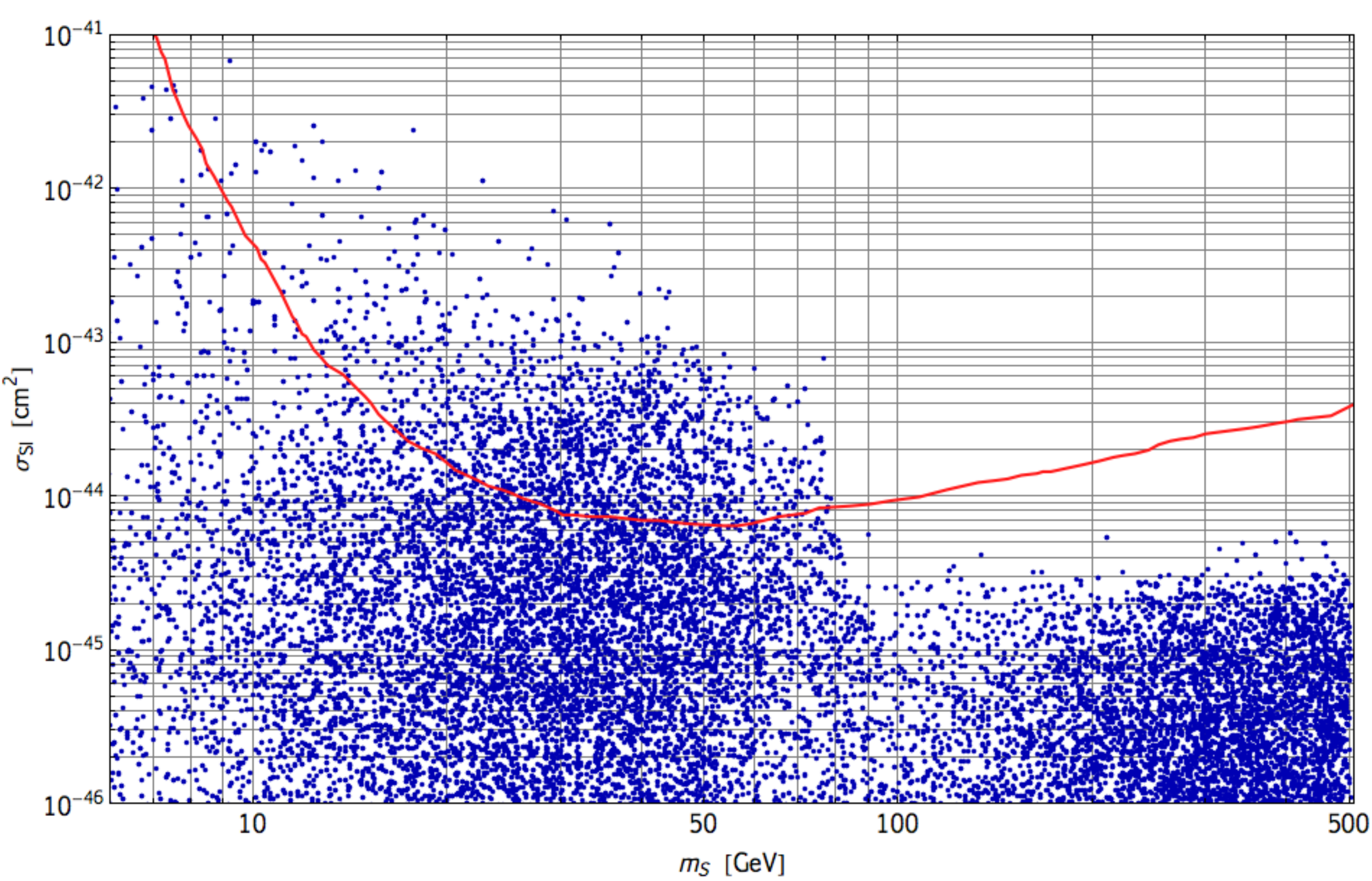}
\caption{Spin-independent cross-section against Dark Matter mass.}
\label{GrapheDMDD}
\end{center}
\end{figure}

The leptogenesis scenario reported above is only possible after the introduction of the scalar $S$, which constitutes a viable Dark Matter candidate. 
Actually, after the electroweak symmetry breaking, the complex $S$ is split into two real fields, the lightest one being the DM. 
First, the stability of this lightest state is insured by a discrete $Z_{2}$ symmetry remnant of the breaking of $U(1)_{B-\tilde{L}}$. 
Second, the relic density of $S$ easily satisfies observational requirements.  
Indeed, the scalar possesses several portal couplings to the $H_{i}$s, couplings which are basically unconstrained although entering in the DM mass: the annihilation cross-section of $S$ is large enough, for a broad range of DM mass. 
Finally, $S$ can scatter on nucleon via those Higgs portal couplings, and a signal in direct detection experiment can be expected in the full DM mass range, as we show in Figure \ref{GrapheDMDD} where we compare the model predictions for the spin-independent DM cross-section (blue dots) with the results of the XENON100 experiment~\cite{XENON100} (red curve). 
\section{Conclusion and Acknowledgments}
In \cite{JosseMichaux:2011ba} we present a UV-completion of the inverse-seesaw model where the baryon asymmetry of the Universe is explained by an original leptogenesis mechanism which arises after the introduction of a scalar singlet field. This latter particle constitutes a viable Dark Matter candidate.\\ \\
This work  is supported by
Funda\c{c}\~{a}o para a Ci\^{e}ncia e a
Tecnologia (FCT, Portugal) through the projects
PTDC/FIS/098188/2008,  CERN/FP/116328/2010
and CFTP-FCT Unit 777,
which are partially funded through POCTI (FEDER).

\end{document}